# A Hybrid Recommendation Method Based on Feature for Offline Book Personalization

Xixi Li[1] *, Jiahao Xing[2], Haihui Wang[2],
Lingfang Zheng[2], Suling Jia[1], Qiang Wang[1]

[1] School of Economic and Management, Beihang University,

Beijing 100089, China

`lixixi199407@buaa.edu.cn, jiasuling@126.com, wang6965@sina.com`

[2] School of Mathematics and System Science, Beihang University,

Beijing 100089, China

`{xingjiahao, whhmath, LINGFANGbeibei}@buaa.edu.cn`

**Abstract.** Recommendation system has been widely used in different areas. Collaborative filtering focuses on rating, ignoring the features of items itself. In order to effectively evaluate customers' preferences on books, taking into consideration of the characteristics of offline book retail, we use LDA model to calculate customers' preference on book topics and use word2vec to calculate customers' preference on book types. When forecasting rating on books, we take two factors into consideration: similarity of customers and correlation between customers and books. Experiment shows that our hybrid recommendation method based on features performances better than single recommendation method in offline book retail data.

## 1   Introduction

Recommender systems were originally defined as ones in which people provide recommendations as inputs, which the system then aggregates and directs to appropriate recipients [1]. Recommender system is aimed at providing personalized goods for customers [2,3,4]. Recommender system has been used in many areas [5,6]. A large amount research has focused on movie recommendation [7,8], music recommendation [9,10], news recommendation [11], hotels recommendation [12], books recommendation [13], e-commerce recommendation [14], tourism recommendation [15] and many other areas. With the development of the recommendation system, many methods and techniques have generated.  The most common method is collaborative filtering [16]. The greatest strength of collaborative techniques is that they are independent of any machine-readable representation of the objects being recommended [17]. Many other techniques have been proposed for performing recommendation, including content-based, knowledge-based, demographic-based and other techniques [17]. To improve recommendation performance, these methods have sometimes been combined in hybrid recommendation system [17].

  Online book recommendation system [18] such as Amazon has been proposed and developed, which brought more profit for retailers and provided personalized service for customers. However it seems that there is a little research about offline book recommendation, which needs to be studied further.  Being different from online book recommendation, online platform can record user behavior (view, like or not, collect, buy and so on) and user generated content [19] with cookie technique, which makes it easier to evaluate customers' preferences on firm's products or services [20]. However, offline book transaction contains sparse transaction information (who buy what book at when), which increases the difficulty of evaluating customers' preference on books and recommending books accurately. Moreover, according to the statistics of our offline book transaction data, the average amount of books that every customer bought is no more than 10, which indicates the sparseness of customer behavior.

  A simple method is that we can build customer-book purchasing matrix according to purchasing behavior with CF (collaborative filtering). To handle the scalability and sparseness problems in CF, several approaches have been developed and the most important one is dimensionality reduction [21], such as SVD [33], LFM [33] and so on. Recommendation systems based on collaborative filtering that make use of ratings to infer hidden product-item features fail for products and items with insufficient number of ratings. Collaborative filtering approach focuses on customer rating data and ignores the features that may attract customers' attention. For example, a customer may like a book because the author, plot theme of the book itself. Therefore, the nature of the item itself is critical to evaluating the customers' preference and thus personalized recommendation. Can we evaluate customers' preference on books from multiple dimensions? The name of books contains comparative richer information, which can be used to extract customers' preference on books. Recent research has explored neural network based to process text data in recommendation methods [19]. Recurrent neural network and convolutional network has been used in this area [34,35]. Representing text using word embedding is shown to improve the representation quality [36]. Can we extract customers' preference on books from books name? Apart from CF, can a hybrid recommender system based on features performance more effectively? Therefore, in order to improve the performance of the recommendation, this paper proposes a hybrid recommendation method based on features. Specifically, we look at two key questions as follows:

- How can we evaluate customer preference from multiple dimensions? How can the knowledge be extracted from book name to evaluate customers' preference on books?
- How the predictive accuracy of hybrid recommendation system can be improved using hybrid recommendation method based on features combined with customer similarity and correlation between customers and books?

---

* Corresponding Author

The remainder of this paper is organized as follows. The related work is explained in Section 2. Section 3 introduces the relevant methods: LDA (Latent Dirichlet Allocation), Word2vec and LFM (Latent Factor Models). Section 4 provides hybrid recommendation methods and experiment result. Conclusions and future work is presented in Section 5.

## 2 Related Work

### 2.1 Recommendation Methods

In this selection, we give a brief overview of methods that have been used in recommender systems and hybrid recommender systems.

A variety of methods have been proposed for recommendation, including collaborative, content-based, knowledge-based, demographic-based and other techniques. Specifically, recommender systems have (1) basic information, the information we get for the recommender system (2) input data, the information we put in the recommender system for processing, and (3) recommendation methods [17]. On this basis, we can distinguish five different recommendation techniques as shown in Tab.1. Assume that I is the set of items over which recommendations might be made, U is the set of users whose preferences are known, u is the user for whom recommendations need to be generated, and i is some item for which we would like to predict u's preference.

Table 1. Different recommender system methods

| Methods | Basic information | Input data | Algorithm |
| --- | --- | --- | --- |
| Collaborative filtering | Ratings from U of items | User's ratings for u of items. | Identify users in U similar to u, and extrapolate from their ratings of i. |
| Content-based | Features of items | User's ratings of items. | Generate a classifier that fits u's rating behavior and use it on i. |
| Knowledge-based | Features of items | A description of user's needs and interests. | Infer a match between i and u's need. |
|  | Knowledge of how these items satisfies a user's needs. |  |  |

Different methods have their own unique features and also their advantage and disadvantages [17]. We can distinguish them as shown in table 2.

Table 2. Tradeoff between recommendation techniques

| Methods | Advantages | Disadvantages |
| --- | --- | --- |
| Collaborative filtering | Identify cross-genre riches; Domain knowledge not needed; Adaptive: quality improves over time; Implicit feedback sufficient | New user ramp-up problem; New item ramp-up problem; "Gray sheep" problem; Quality dependent on large historical data set; Stability vs. plasticity problem |
| Content-based | Domain knowledge not needed; Adaptive: quality improves over time; Implicit feedback sufficient | New user ramp-up problem; New item ramp-up problem; Stability vs. plasticity problem |
| Knowledge-based | No ramp-up required; Sensitive to changes of preference; Can include non-product features | Suggestion ability static; Knowledge engineering required. |

## 2.2 Hybrid Recommendation System

Hybrid recommender systems combine two or more recommendation methods to gain better performance. Most commonly, collaborative filtering is combined with some other technique in an attempt to avoid the ramp-up problem. Table III shows some of the combination methods been employed.

Table 3. Hybridization methods

| Hybridization method | Description |
| --- | --- |
| Weighted | The scores (or votes) of several recommendation techniques are combined together to produce a single recommendation [17]. |
| Switching | The system switches between recommendation techniques depending on the current situation [17]. |
| Mixed | Recommendations from several different recommenders are presented at the same time [17]. |
| Feature combination | Features from different recommendation data sources are thrown together into a single recommendation algorithm [17]. |
| Cascade | Features from different recommendation data sources are thrown together into a single recommendation algorithm [17]. |
| Feature augmentation | Output from one technique is used as an input feature to another [17]. |
| Meta-level | The model learned by one recommender is used as input to another [17]. |

Different hybrid recommendation systems have occurred in different areas. Bash, Hirsh&Cohen [22] put forward a hybrid recommender system combined with CF and CN using feature combination method. CF and KB were combined by Pazzani [23] in 1999 using weighted method. Towle&Quinn [24] proposed a hybrid recommender system of CN and CF by switching method. KB and CF were combined by Resnick [25] using ferture augmentation method.
 (CF=collaborative filtering, CN=content-based, KB=knowledge-based)

# 3 Preliminaries

### 3.1 LDA(Latent Dirichlet Allocation)

(1) The introduction of LDA [26,27]
 The pseudocode of LDA:
   //topic plate
   for all topics $k \in [1, K]$ do
   sample mixture components $\varphi_k \sim Dir(\beta)$
   //document plate:
   For all documents $m \in [1, M]$ do
   sample mixture proportion $\theta_m \sim Dir(\alpha)$
   sample document length $N_m \sim Poiss(\xi)$
   //word plate:
   for all words $n \in [1, N_m]$ in document $m$ do
      sample topic index $z_{m,n} \sim Mult(\theta_m)$
      sample term for word $w_{m,n} \sim Mult(\varphi z_{m,n})$
The schematic diagram below shows the process of generating a document [12]:

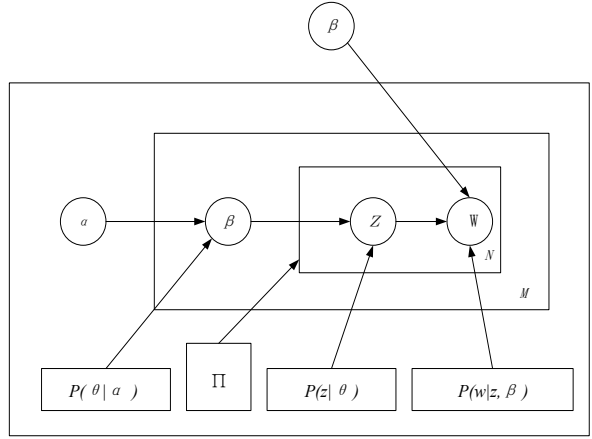

**Fig.1.** Structure of LDA

(2) The indicator of perplexity
   The document $d$'s perplexity in this paper is the training model's uncertainty about which topic the document belongs to. The lower the perplexity is, the better the effect of clustering will be.

$$preplexity(D) = exp\left(-\frac{\sum \log p(w)}{\sum_{d=1}^{M} N_d}\right) \quad (1)$$

Where the denominator is the sum of all the words in the test set, i.e. the length of test set. $p(w)$ means the probability of the occurrence of word $w$ in the test set. The calculation formula is shown as the following:

$$p(w) = p(z|d) \cdot p(w|z) \quad (2)$$

Where $p(z|d)$ represents the probability of topic $z$ appearing in the document $d$. $p(w|z)$ means the probability of word $w$ appearing on the topic $z$.

### 3.2 Word2vec

Shallow distributed representation models are widely used in text processing field, such as word2vec [28,29] and GloVec [30]. Compared with the traditional bag-of-word model, word-embedding model can map words or other information units (for example, phrase, sentences or documents) to a low-dimensional implicit space. In this implicit space, the representation of each information element is a dense eigenvector. The basic idea of word-embedding model is actually from the traditional 'Distributional semantics' [31], which main idea is that the semantics of words are closely related to their adjacent background words. Therefore, this model use embedded representations to build semantic associations between the current and context words.
   There are two important models in Word2Vec: CBOW (Continuous Bag-of-Words) and Skip-gram, which have been introduced in detail in Tomas Mikolov's paper [32].

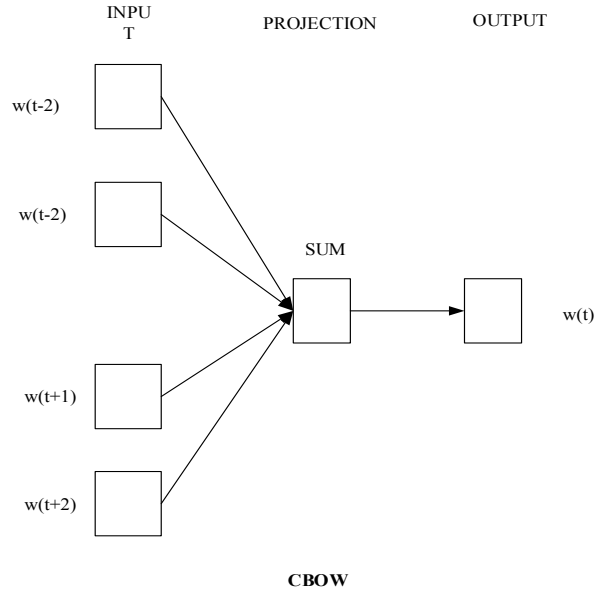

**Fig.2.** Structure of CBOW

$$\sum_{(w,c)\in D} \log P(w|c) \quad (3)$$

$$P(w|c) = \frac{\exp(e\prime(w)^T x)}{\sum_{w\prime \in V} \exp(e\prime(w\prime)^T x)} \quad (4)$$

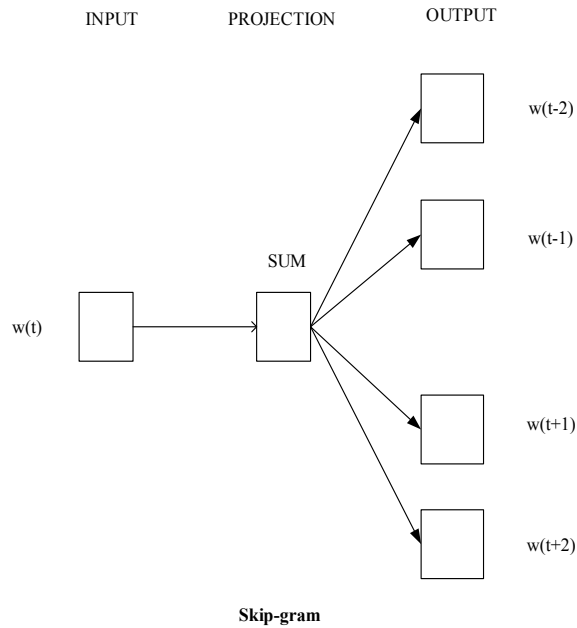

**Fig.3.** Structure of Skip-gram

$$\sum_{(w,c)\in D} \sum_{w_j \in c} \log P(w|w_j) \quad (5)$$

$$P(w|w_j) = \frac{\exp(e\prime(w)^T e(w_j))}{\sum_{w\prime \in V} \exp(e\prime(w\prime)^T e(w_j))} \quad (6)$$

### 3.3 LFM（Latent Factor Model）

Like any other Matrix Factorization approach, SVD model can extract latent feature from the rating matrix. Besides that, SVD [33] is able to simplify data and remove noise. However, plain SVD model also has obvious disadvantages in recommendation system: when customer-item rating comatrix is a sparse matrix (in the real case, it tends to be), the factorization of matrix will need a lot of memory space and increase computational complexity.

Because of the disadvantages of SVD model above, this algorithm has not been widely used in the area of recommendation system. Under these circumstances, Koren put forward a new model based on SVD called LFM (Latent Factor Model).

LFM [33] is another technique of Matrix Factorization. The basic assumption is that there exist an unknown low-dimensional representation of customers and items where customer-item affinity can be modeled accurately. For example, the behavior that a customer buys a book might be assumed to depend on few implicit factors such as the customer's taste across various book themes.

The math principle of LFM is more difficult than SVD. Let $A$ be the initial customer-item matrix, $P$ and $Q$ are matrices broken down from $A$. Some specific elements $a_{ui}$ in matrix $A$ are equal to $p_u^T q_i$. Where A is of size $M \times N$, P is of size $K \times M$ and Q is of size $K \times N$.

Then you can calculate the customer's preference on the item by the following formula:

$$Preference(u, i) = r_{ui} = p_u^T q_i = \sum_{k=1}^{K} p_{uk} q_{ik} \quad (7)$$

In this section, we present the key point of the basic LFM algorithm that is used to create a customer-item similarity matrix. When it comes to recessive feedback problem in recommendation, we need to select negative samples randomly for each customer rather than make all of non-behavior items be zero in the initial matrix. According to other research, it is reasonable to choose unpopular items as negative samples. We will select positive samples and negative samples in the ratio 1 to 10.

Secondly, it is necessary to choose a regularization term to avoid over fitting in this model. The prediction function and the loss function of the model will be:

$$R = (r_{ui})_{m \times n} \approx P \cdot Q^T = (\hat{r}_{ui})_{m \times n} \quad (8)$$

$$\hat{r}_{ui} = \sum_{k=1}^{F} p_{uk} \cdot q_{ik} \quad (9)$$

Loss function:

$$C = \sum_{(u,i)} (r_{ui} - \hat{r}_{ui})^2 = \sum_{(u,i)} (r_{ui} - \sum_{k=1}^{F} p_{uk} \cdot q_{ik})^2 + \lambda \|p_u\|^2 + \lambda \|q_i\|^2 \quad (10)$$

Iterative process:

$$p_{uk} = p_{uk} + \alpha \left(-\frac{\partial C}{\partial p_{uk}}\right) = p_{uk} - \alpha \frac{\partial C}{\partial p_{uk}} \quad (11)$$

$$q_{ik} = q_{ik} + \alpha \left(-\frac{\partial C}{\partial q_{ik}}\right) = q_{ik} - \alpha \frac{\partial C}{\partial q_{ik}} \quad (12)$$

Where $\alpha$ means a learning rate.
$p_{uk}$: the latent factor of the customer $u$;
$q_{ik}$: the latent factor of the item $i$;
$\hat{r}_{ui}$: the correlation between the customer $u$ and the item $i$.

## 4 Experiments

The overall structure of our hybrid recommendation system is as follows:

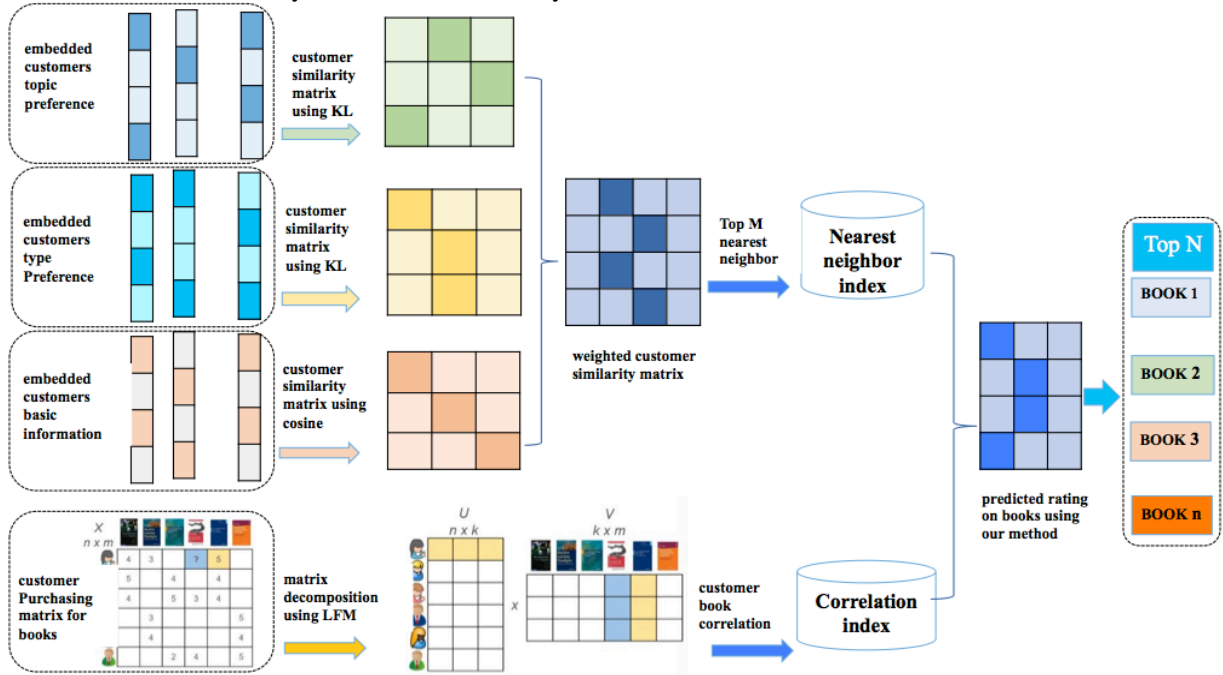

**Fig.4.** Overall structure

The basic information of the data from one offline bookstore is as follow:

**Table 4.** Introduction to the experiment data

| Indicators | Description |
|---|---|
| Timespan | 2016.01.01-2016.12.31 |
| Data Size and Content | 150,000 effective purchasing records and 7,000 customers |

### 4.1 Evaluations of Customers' Preference

(1). Calculating customers' preference on book topics

**Algorithm 1. Calculating customers' preference on book topics from purchasing record**
Input: Purchasing record
Output: Customers preference on book topics
For each Book b
　Begin
　　Calculate the distribution $p_i^{(k)}$ of the book over the topics k using LDA model
　　For each User u
　　　Begin
　　　　Calculate the distribution of customer u on the topic k from their purchasing record
　　　　$p_u^{(k)} = \frac{\sum_{i \in I_u} p_i^{(k)}}{|I_u|}$
　　　End
End

According to algorithm1, we can easily calculate customers' preference on topics of books. We use the name of the book as the document and extract topics by using LDA model. We get the probability distribution $p_i^{(k)}$ of each book i corresponding to the topics $K$. The probability distribution of the customer u over the topic k is:

$$p_u^{(k)} = \frac{\sum_{i \in I_u} p_i^{(k)}}{|I_u|} \tag{13}$$

Where $p_i^{(k)}$ represents the probability that the book i is distributed over the topics k;

$I_u$ represents the collection of books purchased by the customer u;

$|I_u|$ represents the number of books in the collection;

We use the indicator of complexity to determine the number of k. According to complexity, we determine the number of k is 20.

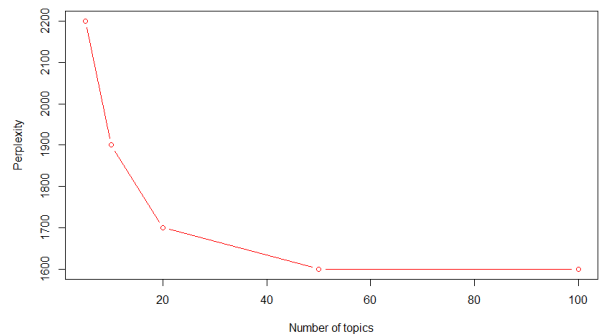

**Fig.5.** Perplexity changes with different number of topics

(2). Calculating customers' preference on book types

**Algorithm 2. Calcaulting customers' preference on book types from purchasing record**
Input: Purchasing record and book type
Output: Customers' preference on book types
For each book type (such as literature, children and so on)
　Begin
　　Calculate the word embedding of book type using word2vec model

```
    End
 For each User u
   Begin
     Calculate the distribution of customer u on the book type from their purchasing record
```
$$c_u^{(m)} = \frac{\sum_{i \in I_u} WordVector_i^{(m)}}{|I_u|}$$
```
   End
```

According to algorithm2, we can calculate customers' preference on type of books. The word embedding of book type is learned by word2vec using the corpus (Chinese Wikipedia's training corpus, the size is about 1G). The dimension k is set as 50. The calculation of the customer's preference on the book type is as follow:

$$c_u^{(m)} = \frac{\sum_{i \in I_u} WordVector_i^{(m)}}{|I_u|} \tag{14}$$

$c_u^{(m)}$: Vector of customer's preference on book type; $m=50$
$I_u$: the collection of books purchased by the customer;
$WordVector_i$: The word vector of the book type, with a dimension of 50.

(3). Vectorizing customers basic information

We can get the customers' age, the way of getting the card, gender and the contact information in the membership management information system and these information is relatively static, which doesn't change with the customers' purchasing behavior. We use one-hot method to vectorize customer basic information.

**Tab.5** Customers' basic information

| Variables | Symbol | Explanation |
|---|---|---|
| Card type | $X_1$ | The way that customer gets card |
| Age | $X_2$ | Discretize age as (0, 20), (20,40), (40,60), (60,100) |
| Gender | $X_3$ | Male, Female |
| Contact | $X_4$ | E.g. message, phone, email contact |

**4.2 Calculation of Similarity between Customers**

(1) Calculation of similarity of topic preference between customers

**Algorithm 3. Calcaulting similarity of book topic preference between customers using KL divergence**
Input: Customer preference on book topics
Output: Distance matrix of customers
For each User $u_1$
  Begin
    For each User $u_2$
      Begin
        Calculate the KL divergence of the two customers:
        $D_{KL}(P||Q) = \sum_i P(i) \cdot \ln \frac{P(i)}{Q(i)}$
        Turn KL divergence symmetrical:
        $D_s(P,Q) = \frac{[D(P,Q)+D(Q,P)]}{2}$
      End
End

The similarity of the distribution can be calculated by the KL divergence. We can use the KL divergence[18] to calculate the similarity of the two customers' preference on the topic of the books.

$$D_{KL}(P||Q) = \sum_i P(i) \cdot \ln \frac{P(i)}{Q(i)} \tag{15}$$

$P(i), Q(i)$: the value of topic preference of customers in topic i.
Since the KL divergence is asymmetric, we turn it symmetrical:

$$D_s(P,Q) = \frac{[D(P,Q)+D(Q,P)]}{2} \tag{16}$$

(2) Calculation of similarity of preference on book type between customers
**Algorithm 4. Calculating similarity of book type preference between customers**
Input: Customers' preference on book type
Output: Distance matrix of customers
For each User $u_1$

```
Begin
   For each User u₂
      Begin
         Calculate the KL divergence of the two customers:
```
$$D_{KL}(P||Q) = \sum_i P(i) \cdot \ln \frac{P(i)}{Q(i)}$$
```
         Turn KL divergence symmetrical:
```
$$D_s(P,Q) = \frac{[D(P,Q)+D(Q,P)]}{2}$$
```
      End
End
```

Here, we use KL divergence to calculate the similarity of the customers' preference on the book type. The specific calculation method is similar to 3.2.1.

In the above, we calculate the similarity of the customers' preference on book topics, type preference and basic information. Next we calculate the similarity between customers by weighted method.

(3) Calculation of weighted similarity between customers

We calculate the similarity between customers by weighted method as follows:

$$sim(u,v) = x \cdot s \tag{17}$$
$$s = (t_{sim(u,v)}, g_{sim(u,v)}, d_{sim(u,v)})$$
$$x_1+x_2+x_3=1$$

$x_1$: weight of similarity of customers' topic preference
$x_2$: weight of similarity of customers' book type preference
$x_3$: weight of similarity of customers' basic information
$t_{sim(u,v)}$ : similarity of customers' topic preference
$g_{sim(u,v)}$: similarity of customers' book type preference
$d_{sim(u,v)}$: similarity of customers' basic information

The distribution of customers' topic preference carries a lot of information reflecting the habits of the customer, so the weight is set to the highest. Three weights are respectively: $x_1$=0.6, $x_2$=0.3, $x_3$=0.1.

### 4.3 Correlation between Customers and Books

Taking into account some problems of SVD and our data itself, we use LFM model to dig out the latent factors between customers and books, and to calculate the correlation between our customers and books.

$$R = (r_{ui})_{m \times n} \approx P \cdot Q^T = (\hat{r}_{ui})_{m \times n} \tag{18}$$
$$\hat{r}_{ui} = \sum_{k=1}^{F} p_{uk} \cdot q_{ik} \tag{19}$$

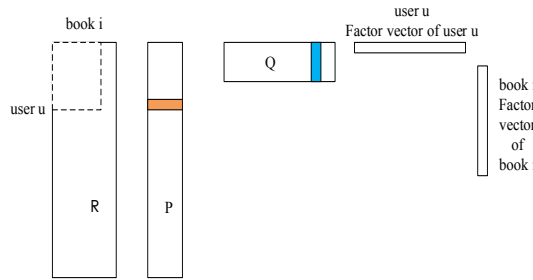

**Fig.6.** Illustration of LFM

(1) Sampling strategy

Compared with online customer behavior, we can use some technology to record various behavior of the customers, for example: clicking record, browsing time, pulling black and so on. While offline book trading system can only record the behavior of customers that buy or not buy, which determines our customer-book matrix is sparse. We put the number 1 in the corresponding position of matrix if the customer has a purchasing behavior and record 0 if not.

The problem is that the customer does not buy a book does not mean that customers really do not like that book. When we deal with the behavior that the customer does not buy the book, we use a sampling method to determine our positive and negative samples. According to the number of books purchased as the book pool, we use the following strategy to extract negative samples for each customer:
  1) The number of positive and negative samples remains the same
  2) The probability of the item being drawn is proportional to the popularity of the object

(2) Algorithm optimization
1) Comparison of that negative samples are fixed and randomized

**Fig.7.** Comparison of that negative samples are fixed and randomized

It can be seen from the figure that the fixed negative samples in the iterative process are more likely to obtain better solutions than the random negative samples.
2) Comparison of alpha

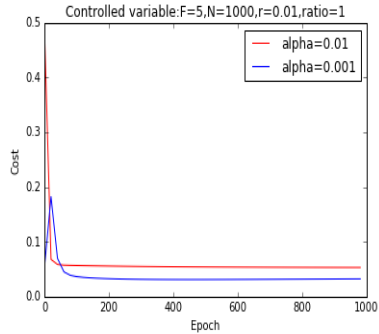

**Fig.8.** Comparison of alpha

From the figure, the smaller alpha behaves slightly better than the larger alpha value when the iteration process is larger (> 100).
3) Comparison of proportions of positive and negative samples

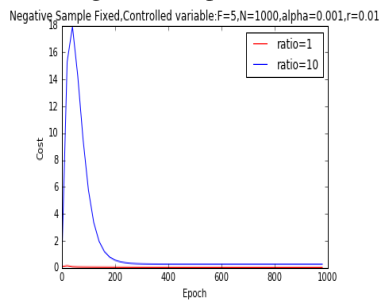

**Fig.9.** Comparison of proportions of positive and negative samples

The figure shows that the larger the proportion of positive and negative samples is, the greater the loss function is in the early stage, and the optimization effect changes greatly with the iterative process, and the final optimization effect is not as good as the same proportions of positive and negative samples.
4) Latent factors and regularization parameters

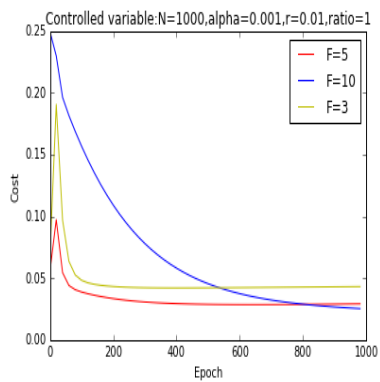

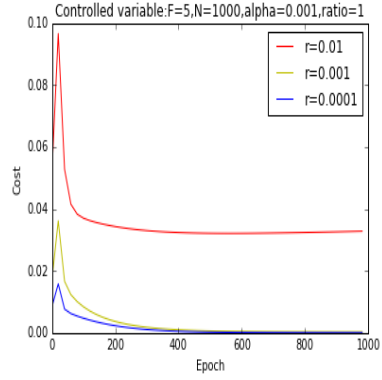

**Fig.10.** Latent factors and regularization parameters

The two figures show that the number of latent factor and the regularization coefficient have little effect on the optimization effect when the iterative process is large enough.

By adjusting the different parameters, we take:
(1) Fixed negative samples
(2) Take adaptive learning rate
(3) The ratio of negative samples is 1
(4) The number of latent factors is 5
(5) The regularization coefficient is 0.01

Through LFM model, we can get the latent factors of customer and book and calculate the correlation $\hat{r}_{ui}$ between the customer and the book.

### 4.4 Calculation of Predicted Rating

**Algorithm 5. Calculating the predicted rating**
Input: Similarity matrix of customers and correction between customers and book
Output: Predicted rating on book for each customer
For each Customer u
  Begin
    Find the N (N=10,12,14,16,18,20,22,24,26,28) neighborhood for customer u
    Let $w_i$(weight of neighborhood similarity)=1/(N+1)
    Let $w_2$(weight of correction with book of customer themself)=1-N/(N+1)
    Calculate predicted rating on book:
    $p(u,i) = \sum_{v \in N_u, i \in I}(w_i \cdot sim(u,v) \cdot \hat{r}_{vi} + w_2 \cdot \hat{r}_{ui})$
  End

We calculate the similarity between customers, for each customer u we can find the customer u neighborhood N, predict the customer's rating on book i p (u, i):

$$p(u,i) = \sum_{v \in N_u, i \in I}(w_i \cdot sim(u,v) \cdot \hat{r}_{vi} + w_2 \cdot \hat{r}_{ui}) \quad (20)$$

$$\sum_{i \in I} w_i + w_2 = 1 \quad (21)$$

$\hat{r}_{vi}, \hat{r}_{ui}$ respectively represent the correlation between customer v, u and book $i$

$w_i, w_2$ represent the weight of the similarity between the customers and the weight of the target customer's $u$ correction to the book $i$.

Here, we discuss how to select the weight, we set the weight coefficient as follows:

$$\sum_{i \in I} w_i = \frac{n}{n+1} \quad (22)$$

$$w_2 = 1 - \frac{n}{n+1} \quad (23)$$

$n$ represents the number of *TopN* customers close to the customer $u$, we choose $n= 10$

According to the ranking of predicted rating on books, we select top N (N=10,20) books for recommendation.

## 5 Evaluation of recommendation system

### 5.1 Experimental metric

Precision is a metric that represents the probability that an item recommended as relevant is truly relevant. It is defined as the ratio of items correctly predicted as relevant among all the items selected:

$$precesion = \frac{\sum_{u \epsilon U} |R(u) \cap T(u)|}{\sum_{u \epsilon U} |R(u)|} \tag{24}$$

Recall is a metric that represents the probability that a relevant item will be recommended as relevant. It is defined as the ratio of items correctly predicted as relevant among all the items known to be relevant:

$$recall = \frac{\sum_{u \epsilon U} |R(u) \cap T(u)|}{\sum_{u \epsilon U} |T(u)|} \tag{25}$$

F is the weighted average of precision and recall.

$$F = \frac{precesion * recall * 2}{precesion + recall} \tag{26}$$

R(u): items recommended for customers

T(u): items purchased by customers

### 5.2 Experimental results

When we use single recommendation methods LFM, the average precision is about 7% with the change of number of the recommended books. However, when we use hybrid recommendation system, the precision is approximately 20%, which is greatly improved. In the area of offline book, our hybrid method is meaningful.

Impact of size of TopN in different neighbors

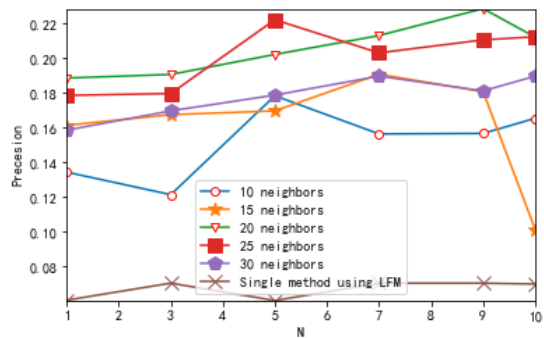

**Fig.11.** Precision in different numbers of recommended books

When we use single recommendation methods LFM, the average recall is about 8% with the change of number of recommended books. However, when we use hybrid recommendation system, the precision is approximately 25%, which is greatly improved. In the area of offline book, our hybrid method is meaningful.

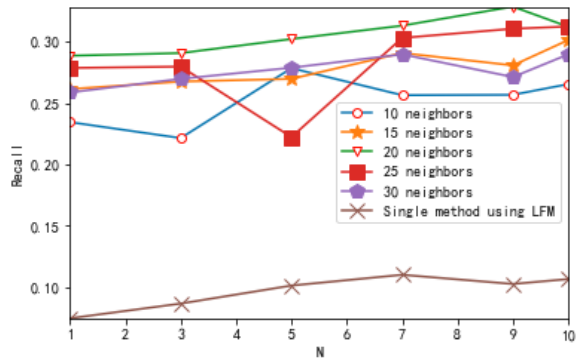

**Fig.11.** Recall in different numbers of recommended books

When we use single recommendation methods LFM, the average F is about 7.5% with the change of number of recommended books. However, when we use hybrid recommendation system, the precision is approximately 24%, which is greatly improved. In the area of offline book, our hybrid method is meaningful.

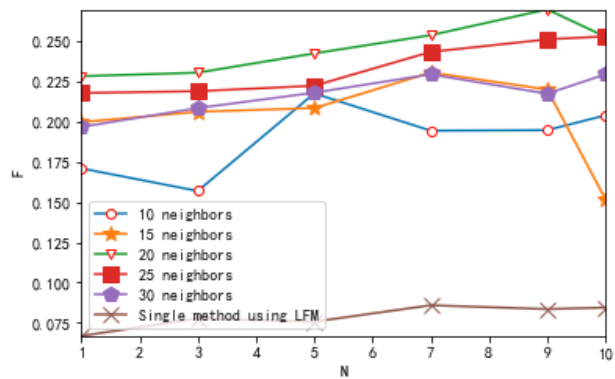

**Fig.11.** F in different numbers of recommended books

## 6   Conclusion

Recommendation system has been widely used in different areas. Collaborative filtering focuses on rating, ignoring the features of items itself. In order to better evaluate customer preference on books, we use LDA model to calculate customer preference on book topics and use word2vec to calculate customer preference on book types. In order to forecast rating on books, we take two factors into consideration: similarity of customers and correlation between customers and books. Experiment shows that our hybrid recommendation method based on features performances better in offline bookstore data.

What we contribute is to solve two key problems: (1) we evaluate customer preference from multiple dimensions. (2) combined with the characteristic of offline book transaction, we proposed a new hybrid recommendation method based on features to improve the performance.

However, this paper has shortcomings as follows: we take a qualitative approach to set the weight in the calculation of weighted similarity between customers, which needs to be further improved; we use the average method to set the weight to adjust the predicted rating in this article, whose rationality needs to be further improved. In addition, we use word2vec to calculate customer preference on book types, the dimension is selected as 50, which needs further discussion. Moreover, our method has yet to be tested on other data sets for its performance.

## References

[1] Resnick P, Varian H R, Special issue on recommender systems, Ai Communications 21 (2-3)(1997) 95-96.

[2] Jannach, D., Gedikli, F., Karakaya, Z., Juwig, O., Recommending hotels based on multi-dimensional customer ratings, In Information and communication technologies in tourism (2012) 320–331.

[3] Jannach, D., Finding preferred query relaxations in content-based recommenders, in: Proc. In Third international IEEE conference on intelligent systems, 2006.

[4] Jannach, D., Karakaya, Z., Gedikli, F., Accuracy improvements for multi-criteria recommender systems, in: Proc. the 13th ACM conference on electronic commerce, 2012.

[5] Park, D. H., Kim, H. K., Choi, I. Y., Kim, J. K., A literature review and classification of recommender systems research, Expert Systems with Applications 39(2012a) 10059–10072.

[6] Park, D. H., Kim, H. K., Choi, I. Y., Kim, J. K., A literature review and classification of recommender systems research, Expert Systems with Applications 39(2012b) 10059–10072.

[7] Anand, D., Mampilli, B. S., Folksonomy-based fuzzy user profiling for improved recommendations, Expert Systems with Applications 41(2014) 2424–2436.

[8] Carrer-Neto, W., Hernández-Alcaraz, M. L., Valencia-García, R., García-Sánchez, F., Social knowledge-based recommender system, Application to the movies domain, Expert Systems with Applications 39(2012) 10990–11000.

[9] Bogdanov, D., Haro, M., Fuhrmann, F., Xambó, A., Gómez, E., Herrera, P., Semantic audio content-based music recommendation and visualization based on user preference examples, Information Processing & Management 49(2013) 13–33.

[10] Hyung, Z., Lee, K., Lee, K., Music recommendation using text analysis on song requests to radio stations, Expert Systems with Applications 41(2014) 2608–2618.

[11] Das, A. S., Datar, M., Garg, A., Rajaram, S., Google news personalization: Scalable online collaborative filtering, in: Proc. the 16th international conference on World Wide Web (2007) 271–280.

[12] Jannach, D., Gedikli, F., Karakaya, Z., Juwig, O., Recommending hotels based on multi-dimensional customer ratings, In Information and communication technologies in tourism (2012) 320–331.

[13] Núñez-Valdéz, E. R., Cueva Lovelle, J. M., Sanjuán Martínez, O., García-Díaz, V., Ordoñez de Pablos, P., Montenegro Marín, C. E., Implicit feedback techniques on recommender systems applied to electronic books, Computers in Human Behavior 28(2012) 1186–1193.

[14] Castro-Schez, J. J., Miguel, R., Vallejo, D., López-López, L. M., A highly adaptive recommender system based on fuzzy logic for B2C e-commerce portals, Expert Systems with Applications 38 (2011) 2441–2454.

[15] Fuchs, M., Zanker, M, Multi-criteria ratings for recommender systems: An empirical analysis in the tourism domain, In E-commerce and web technologies (2012) 100–111.

[16] Brodt T, Collaborative Filtering, Computer Science 57(4)(2002) 189-189.

[17] Burke, Robin, Hybrid Recommender Systems: Survey and Experiments, User Modeling and User-Adapted Interaction 12(4)(2002) 331-370.

[18] Smith B, Linden G, Two Decades of Recommender Systems at Amazon.com, IEEE Educational Activities Department, 2017.

[19] Chelliah M, Sarkar S., Product Recommendations Enhanced with Reviews, Eleventh ACM Conference on Recommender Systems (2017) 398-399.

[20] Kuhfeld, W. F., Marketing research methods in SAS, In Marketing research methods in the SAS system: A collection of papers and handouts (2005) 21–46.

[21] Cho, Y. H., Kim, J. K., Application of Web usage mining and product taxonomy to collaborative recommendations in e-commerce, Expert Systems with Applications 26 (2004) 233–246.

[22] Basu, C., Hirsh, H. and Cohen W., Recommendation as Classification: Using Social and Content-Based Information in Recommendation, in: Proc. the 15th National Conference on Artificial Intelligence, Madison WI (1999) 714-720.

[23] Billsus, D., Pazzani, M., User Modeling for Adaptive News Access, User-Modeling and User-Adapted Interaction (2000) 147-180.

[24] Towle, B., Quinn, C.: 2000, Knowledge Based Recommender Systems Using Explicit User Models, In Knowledge-Based Electronic Markets, Papers from the AAAI Workshop, AAAI Technical Report WS-00-04, (2000) 74-77.

[25] Resnick, P., Iacovou, N., Suchak, M., Bergstrom, P. and Riedl, J., GroupLens: An Open Architecture for Collaborative Filtering of Netnews, in: Proc. the Conference on Computer Supported Cooperative Work, Chapel Hill, NC, (1994) 175-186.

[26] Blei D M, Ng A Y, Jordan M I, Latent dirichlet allocation, JMLR.org, 2003.

[27] Wang X, Wang Y, Improving Content-based and Hybrid Music Recommendation using Deep Learning, ACM International Conference on Multimedia, (2014) 627-636.

[28] Baxla M A, Comparative study of similarity measures for item based top n recommendation, 2014.

[29] Elbadrawy A, Karypis G, User-Specific Feature-Based Similarity Models for Top-n Recommendation of New Items, Acm Transactions on Intelligent Systems & Technology 6(3)(2015) 1-20.

[30] Wei X, Croft W B, LDA-based document models for ad-hoc retrieval, (2006) 178-185.

[31] Daoud Clarke, A Context-Theoretic Framework for Compositionality in Distributional Semantics, Computational Linguistics 38(1)(2012) 41-71.

[32] Mikolov T, Chen K, Corrado G, et al, Efficient Estimation of Word Representations in Vector Space, Computer Science, 2013.

[33] Koren Y, Bell R, Volinsky C, Matrix Factorization Techniques for Recommender Systems, IEEE Computer Socie-ty Press, 2009.

[34] Elkahky, A.M., Song, Y., and He, X., A multi-view deep learning approach for cross domain user modeling in recommendation systems, in: Proc. the 24th International Conference on World Wide Web, 278-288, 2015.

[35] Cao, L, Non-IID Recommender Systems: A Review and Framework of Recommendation Paradigm Shifting, Engineering 2(2)(2016) 212-224.

[36] Mikolov T, Sutskever I, Chen K, et al, Distributed Representations of Words and Phrases and their Compositionality (2013) 26:3111-3119.